\documentclass[letter]{aa} 
\usepackage{graphicx}
\usepackage{txfonts}
\begin{document}
   \title{Low energy H+CO scattering revisited:} 
    \subtitle{CO rotational 
          excitation with new potential surfaces}

   \author{B. C. Shepler,\inst{1} B. H. Yang,\inst{2} T. J. Dhilip Kumar,\inst{3}
           P. C. Stancil,\inst{2} J. M. Bowman,\inst{1} N. Balakrishnan,\inst{3}
           P. Zhang,\inst{4} E. Bodo,\inst{4,5}
           \and
           A. Dalgarno\inst{4}
          }

   \institute{
             Department of Chemistry and Cherry L. Emerson Center for
             Scientific Computation, Emory University, Atlanta, GA 30322, USA\\
             \email{bcshepl@, jmbowman@emory.edu}
         \and
              Department of Physics and Astronomy and the
              Center for Simulational Physics, The University of
              Georgia, Athens, GA, 30602-2451, USA.
              \email{yang@, stancil@physast.uga.edu}
         \and
              Department of Chemistry, University of Nevada Las Vegas,
              Las Vegas, NV 89154, USA\\
              \email{dhilip.thogluva@unlv.edu, naduvala@unlv.nevada.edu}
         \and
              Harvard-Smithsonian Center for Astrophysics,
              60 Garden St., Cambridge, MA 02138, USA\\
              \email{pezhang@, adalgarno@cfa.harvard.edu}
         \and
              Department of Chemistry and CNISM, University of Rome
              ``Sapienza", Piazzale, A. Moro, 00185, Rome, Italy\\
              \email{e.bodo@caspur.it}   
           }

   \date{Received XX 2007; accepted XX 2007 }

 
  \abstract
   {A recent modeling study of brightness ratios for CO rotational transitions in
    gas typical of the diffuse ISM by Liszt found the role of H collisions to
    be more important than previously assumed. This conclusion 
    was based on recent quantum scattering calculations using the so-called WKS
    potential energy surface (PES) which reported a large cross section for
    the important $0\rightarrow 1$ rotational transition. This result is in contradiction to
    one obtained using the earlier ``BBH" PES for which the cross section is quite
    small and which is consistent with an expected homonuclear-like 
    propensity for even $\Delta J$ transitions.}
   {To revisit this contradication with new scattering 
    calculations using two new ab initio PESs
    that focus on the important long-range behavior and to explore the 
    validity of the apparent departure from the expected 
   even $\Delta J$ propensity
    in H-CO rotational excitation obtained with the WKS PES.}
   {Close-coupling (CC) rigid-rotor calculations for CO($v=0,J=0$) excitation
    by H are performed on four different PESs. Two of the PESs are obtained
    in this work using state-of-the-art quantum chemistry techniques at the
    CCSD(T) and MRCI levels of theory.} 
   {Cross sections for the $J=0\rightarrow 1$, as well as other odd $\Delta J$,
    transitions are significantly suppressed compared to even $\Delta J$
    transitions  in thermal energy CC calculations using the CCSD(T) and MRCI
    surfaces. This is consistent with the expected even $\Delta J$ propensity
    and in contrast to CC calculations using the WKS PES which predict a
    dominating $0\rightarrow 1$ transition.}
   {Inelastic collision cross section calculations are sensitive to fine
    details in the anisotropic components of the PES and its long-range      
    behavior. The current results
    obtained with new surfaces for H-CO scattering suggest that the original
    astrophysical assumption that excitation of CO by H$_2$ dominates the kinetics
    of CO in diffuse ISM gas is likely to remain valid.}

   \keywords{molecular processes --
                ISM: molecules 
               }
   \authorrunning{Shepler et al.}
   \titlerunning{H+CO scattering revisted}
   \maketitle
%

\section{Introduction}

Carbon monoxide is found in a variety of astrophysical environments
usually being the most abundant molecule with the exception of H$_2$.
It is typically observed in absorption in the UV and near-IR and in emission in
the far-IR to millimeter wavelengths. For example,
the {\it Infrared Space Observatory (ISO)}
detected a large number CO rovibrational lines toward Orion Peak 1 and 2
(\cite{gon02}) while pure rotational lines were observed in the planetary
nebulae NGC 7027 (\cite{cer97}).

In most environments, the molecular
level populations are not in equilibrium, so that predicting spectral
line intensities depends crucially on the magnitude of collisional 
excitation rate coefficients due to the dominant species H$_2$, He, and H.  
Adopting semiempirical potentials, Green \& Thaddeus (1976) performed
one of the earliest quantal scattering calculations for the CO-H$_2$, CO-He,
and CO-H systems at astrophysically relevant temperatures. Their results
suggested that the rate coefficients for atomic hydrogen collisions were
significantly less than those due to H$_2$, so that CO-H collisions were
usually neglected in modeling studies of CO spectra.                     

The H-CO complex has been extensively studied by the chemical physics community
for the past three decades including early ab initio PES calculations by
Botschwinna (1974). A later ab initio PES by Bowman, Bittman, and Harding
(1986, BBH) was constructed and used in subsequent scattering calculations
(\cite{lee87}). The H+CO system has a barrier to the formation of the HCO complex
at a H-CO internuclear separation of 4.2 a$_0$ with a calculated height
of 0.125-0.17 eV, but measured to be
0.087(17) eV ($\sim$700 cm$^{-1}$ or 1000 K) (\cite{wan73}).
Lee \& Bowman (1987) found that for collision energies below the barrier,
and not coincident with tunneling resonances, rotational excitation cross
sections displayed an even $\Delta J$ propensity, i.e. homonuclear-like
behavior at long-range. 

Another H-CO surface was computed at the MRCI level of theory by Keller
et al. (1996), the so-called WKS PES. This PES was found to reproduce H-CO
spectroscopic measurements better than that of the BBH surface. It was also
noted by Green et al. (1996) that the WKS and BBH surfaces gave
different rotational excitation cross sections with the difference being
most significant for the $0\rightarrow 1$ transition. This discrepancy was
ascribed to the differences in the treatment of the long-range portion of
the potentials. Balakrishnan, Yan, \& Dalgarno (2002) later
confirmed this difference in the rotational excitation cross sections
and therefore utilized the WKS surface for large-scale calculations of 
rotational excitation rate coefficients for astrophysical modeling
applications.

Recently, Liszt (2006) adopted the new CO-H rotational excitation rate
coefficients of Balakrishnan et al. (2002) in an investigation of the
non-equilibrium rotational populations of CO in environments typical
of the diffuse interstellar medium (ISM). He concluded that the increase
in the rotational excitation rate coefficients for CO-H substantially
enhanced CO rotational brightness and excitation temperatures and that
CO-H collisions should not be neglected in future models.

However, what has been previously overlooked is that the new large $0\rightarrow 1$ 
rate coefficient results in a departure from the homonuclear-like even $\Delta J$
propensity which has been argued on reasonable physical grounds to
be valid for the H-CO collision system (\cite{lee87}). As the scattering results have
been confirmed in a number of theoretical investigations, 
a possible explanation is a subtle deficiency in the long-range portion of the 
WKS surface. In this letter, we explore this issue by computing two new and
completely independent H-CO rigid-rotor surfaces and performing close-coupling 
calculations to obtain the rotational excitation cross sections. We briefly describe
the potential surface computations and scattering calculation methods in the
next two sections. The results and their astrophysical implications are
presented in Section 4 followed by a conclusion in Section 5.


\section{Potential Energy Surface Calculations}

One of the two new ab initio surfaces was calculated using the 
coupled cluster method with single and double excitations and a 
perturbative treatment of triple excitations [CCSD(T)] employing the 
frozen core approximation 
(Purvis \& Bartlett 1982,
 Raghavachari et al. 1989). For the open-shell calculations, 
restricted open-shell Hartree-Fock orbitals were used in the CCSD(T) calculations, 
but the spin-restrictions were relaxed in the solution of the coupled cluster 
equations [R/UCCSD(T)] 
(Knowles et al. 1994).
We adopted the doubly augmented correlation consistent d-aug-cc-pVnZ ($n$ = T, 
Q, and 5) basis set of Woon \& Dunning (1994). The full counterpoise correction was 
applied to approximately account for basis set superposition error (Boys \& Bernardi 
1970).  At each configuration, calculations were performed with all three basis 
sets for the dimer (HCO) and monomers (CO and H) and these energies were 
extrapolated to the complete basis set (CBS) limit. These extrapolated total
energies were then used to compute the 
CBS limit counterpoise corrected interaction energies.  
The interaction potential was computed 
in Jacobi coordinates with the CO bond length $r$ fixed at its experimental 
equilibrium value (2.1322 a$_0$) (Huber \& Herzberg 1979).  The resulting 
rigid-rotor PES that is a function of $R$ (distance between H and the CO center 
of mass) and $\theta$ (angle between the vectors ${\bf r}$ and 
${\bf R}$) was fit to the expression
\begin{equation}
V(R,\theta) = \sum_{\lambda=0}^N P_\lambda (\cos \theta ) V_\lambda (R)
\end{equation}
where the $P_\lambda$ are normalized Legendre polynomials with  $N = 12$.  The 
$V_\lambda (R)$ were calculated using 12th-order Gauss-Legendre quadrature.
The two-dimensional grid on which the ab initio calculations were carried out 
corresponded to 23 values of $R$ between 3-20 a$_0$ and the 12 values of $\theta$
 were obtained from the 12 Gauss-Legendre quadrature points.  The large range 
in $R$ was chosen in order to obtain good coverage of the long-range interaction, 
which is important in the low-energy scattering, as discussed below.  The 
calculations were carried out with the MOLPRO suite of ab initio programs 
(\cite{amo02}).

The second PES was computed using the complete active space self-consistent field  
[CASSCF] (\cite{wk_jcp82,kw_cpl115}) and internally contracted multireference
configuration interaction [MRCI] (\cite{kw_cpl145, kw_tca84, wk_jcp89})
method with aug-cc-pVQZ basis set for H, C, and O by Woon \& Dunning (\cite{woo94})  
as implemented in the 
MOLPRO program.    
Calculations were performed for 1-3~$^1$A$'$ states in the $C_s$ point group for
all non-linear approaches in Jacobi coordinates, and for a CO bond length of
$r=2.20$~a$_0$.
The rigid rotor PES was obtained for the following grid of points:
$R=$ 1.6-5.0 (0.2), 5.0-5.8 (0.4), and 6.0-15.0~a$_0$ (1.0~a$_0$) and $\theta$=1-179$^\circ$
(15$^\circ$), where the numbers in parentheses indicate the step-sizes for the
corresponding range and coordinate. The final energies incorporate the
Davidson correction (\cite{ld_ijqc8}). The computed potential energy values were 
combined with an accurate  ab initio long-range potential that      
includes $C_6$, C$_7$, and C$_8$ coefficients for the H-CO interaction.
These coefficients have been obtained using Pad\'e approximants with
Cauchy moments evaluated by linear response coupled cluster (LR-CCSD)
theory (H\"attig et al. 1997) with the d-aug-cc-pwCV5Z basis set 
(Peterson \& Dunning 2002) and
correlating all 14 electrons of CO in the CCSD calculation. The long
range part has been smoothly
joined to the ab-initio calculated points and 
a full analytical representation of the rigid rotor potential with accurate
long-range interaction was constructed and used in the scattering calculations.

\section{Collisional Excitation Calculations}

In the present study, 
the quantum close-coupling (CC) method 
(Arthurs \& Dalgarno 1960, Lester 1971, Secrest 1979) was used for
the calculations of rotational excitation due to H atom impact.
CO was treated in the rigid-rotor approximation  being held
fixed at its equilibrium bond-length.
The CC scattering
calculations were performed by expanding the angular dependence of the
relevant potential in Legendre polynomials as given in Eq. (1).
Calculations were performed with the non-reactive molecular scattering progam MOLSCAT
(\cite{molscat}) to generate integral state-to-state cross sections.
The CC equations were integrated using the modified
log-derivative Airy propagator of Alexander and Manolopoulos (\cite{Alexander})
with a variable step-size.
A sufficient number of total angular momentum partial waves
and basis set terms have been included to secure convergence of the cross sections.

\section{Results and Discussion}

Figure 1a compares the first four terms ($\lambda=0-3$) of the Legendre expansion for
the BBH and WKS surfaces as well as the new PESs computed in the current
work, referred to as CCSD(T) and MRCI. The agreement between all surfaces
appears to be reasonable at the scale of the figure.
However, focusing on the long-range portion, clear differences are apparent 
in the $\lambda=1$ component as shown in Fig. 1b.
The present PESs are believed to be more reliable than previous surfaces
because we have specifically focused on the long-range behavior through
explicit calculations at much larger internuclear distances than in the
earlier work. Further, the importance of the long-range behavior and
its influence on low-energy collisions is elevated due to the barrier 
at $R=4.2$ a$_0$ as described below.   


   \begin{figure*}
   \centering
   \includegraphics[width=13.0cm,clip]{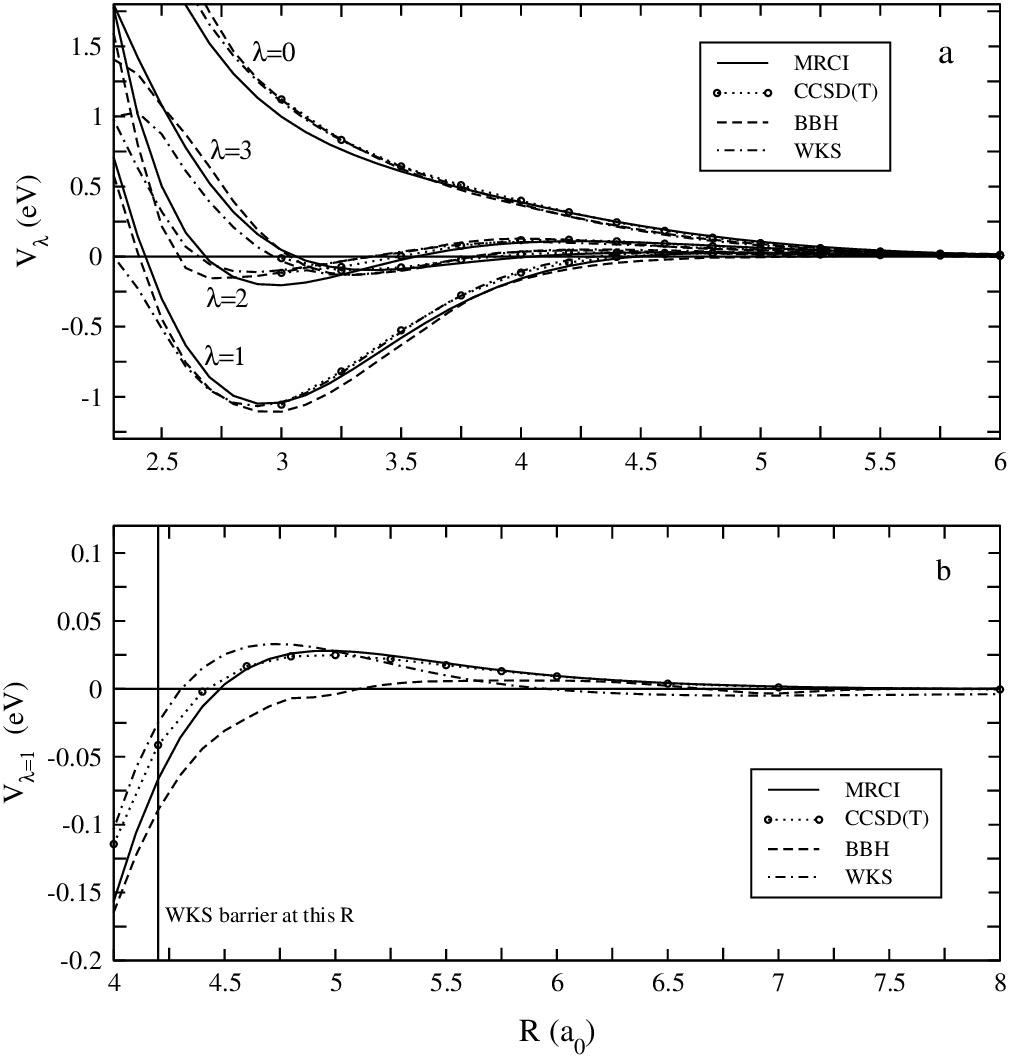}
      \caption{(a) Legendre expansion terms $V_{\lambda}(R)$ for the 
              H-CO PES with the CO internuclear distance $r$ fixed
              at its equilibrium value. (b) Same as in (a), but focusing
              on the long-range behavior for $\lambda=1$. See text for
              descriptions of the different surfaces.}
         \label{Fig1}
   \end{figure*}
%

Calculations of state-to-state cross sections from initial rotational state $J=0$
for the excitation of CO by H on the MRCI, CCSD(T), BBH, and  WKS surfaces
were performed for collision energies at 400 and 800 cm$^{-1}$ as given in
Figs. 2 and 3, respectively. The cross sections using the BBH and WKS surfaces
are in good agreement with those obtained by Green et al. (1995) and 
Balakrishnan et al. (2002), respectively. While the results presented here
adopted the rigid-rotor approximation, we performed tests with BBH and
WKS and found that the resulting cross sections using the full
surfaces, i.e., with vibrational motion, are practically identical.
The largest difference was less than $\sim$10\% for the $0\rightarrow 1$
transition with WKS. Our rigid-rotor cross sections in Figs. 2 and 3 further
confirm the previously noted result that the $0\rightarrow 1$ transition
using WKS is approximately an order of magnitude larger than that obtained with BBH.

Comparing now the rigid-rotor cross sections obtained with the new CCSD(T) and
MRCI surfaces, we see that i) the two sets of cross sections are in very
good agreement,
ii) they reproduce the even $\Delta J$ propensity obtained
with BBH, and iii) the $0\rightarrow 1$ transition is small with a value
similar to that obtained with BBH, in contrast to the WKS result.
These observations can be interpreted by examining the $\lambda=1$ terms
displayed in Fig. 1b since the dominant contributions to the state-to-state
cross sections originate from potential coupling matrix elements with
 $\lambda=|\Delta J|$. The $\lambda=1$ components from the CCSD(T) and MRCI
calculations are seen to be very similar, both displaying a peak near 5 a$_0$
and comparable long-range behavior. Conversely, for the WKS surface the
$\lambda=1$ component peaks at a shorter internuclear distance, closer to the
barrier location, and has a different long-range behavior. 
Indeed, the long-range tail of the 
WKS potential appears to be decaying slower than the 
other potentials. The BBH
$\lambda=1$ component is smaller than that of the other surfaces which
may explain its slightly smaller $0\rightarrow 1$ cross section.

To further test the apparent discrepancy related to the $\lambda=1$ component
of the Legendre expansion, we examined scattering results for  excitation
of the $J=1$ and 2 rotational states of CO($v=0$). The cross sections for
the $\Delta J$=1 excitation transitions with the WKS surface were found to
also be an order of magnitude larger than those obtained with the other
surfaces. 

Based on all of our scattering calculations in conjuction with
the computation of the new surfaces, it appears that the WKS surface has
a different anisotropic behavior, especially at long-range. This anistropy, evident in the
$\lambda=1$ component, while of small magnitude, drives the $\Delta J=1$
transitions and results in the significant discrepancy in the important
$J=0\rightarrow 1$ cross section. 

Our findings have important implications for modeling of CO in non-equilibrium
situations, e.g. diffuse interstellar gas, where atomic hydrogen is present.
The conclusions of Liszt (2006), based on the adoption of the rotational
excitation rate coefficients obtained with WKS, that H collisions can
significantly affect the CO rotational populations, and therefore modeling of
the resulting absorption/emission lines, is no longer valid. In particular,
the $0\rightarrow 1$ cross section due to H impact  is found to be much 
smaller than that due to H$_2$ collisions in agreement with the earlier
results of Chu \& Dalgarno (1975) and Green \& Thaddeus (1976). Further,
our results illustrate the sensitive dependence that low-energy scattering
can have on fine details of the potential energy surface.

   \begin{figure}
   \centering
   \includegraphics[width=8.0cm,clip]{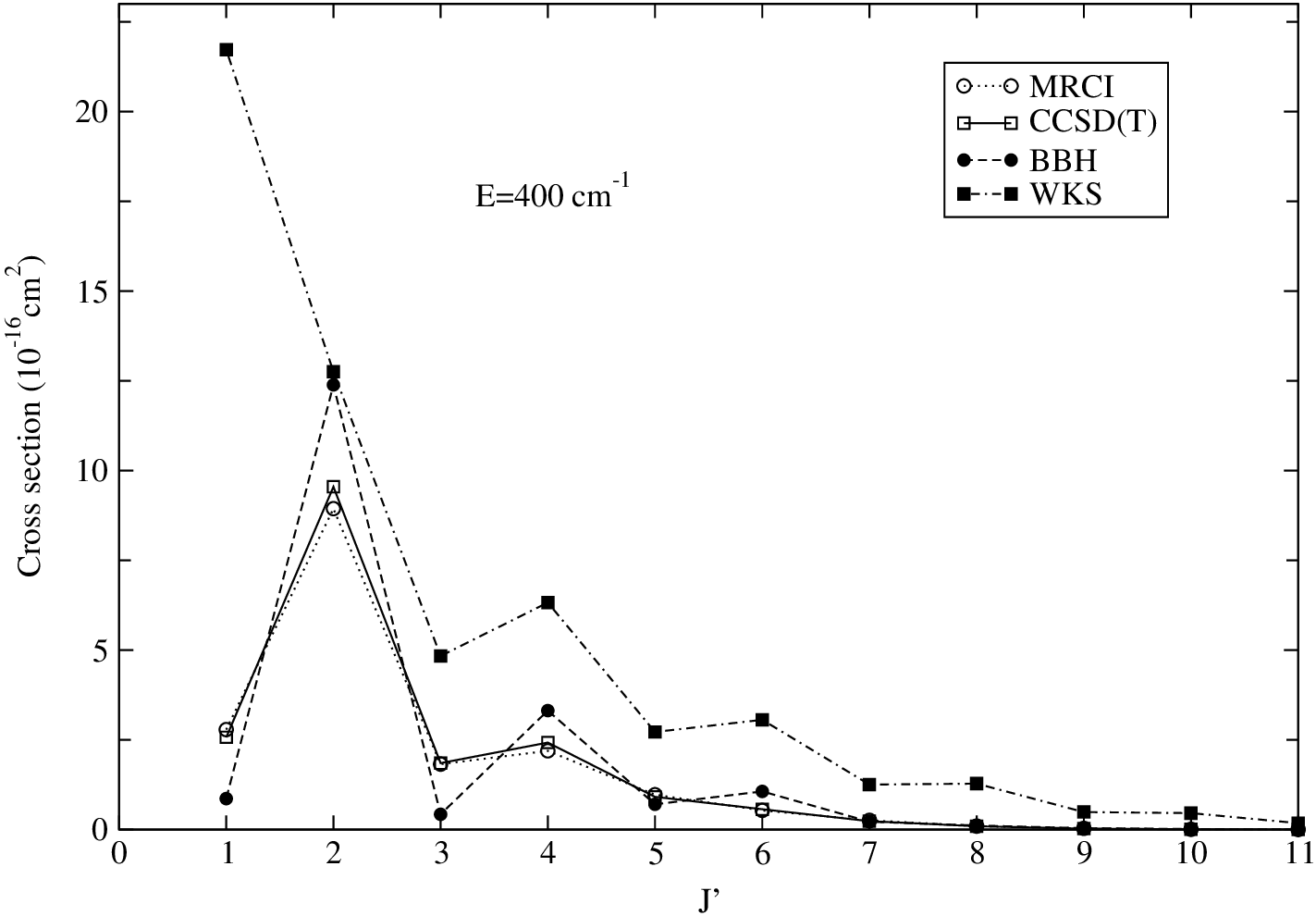}
      \caption{Excitation cross sections due to H impact for
       CO($v=0,J=0$) to CO($v'=0,J'$) for a center of mass collision energy
       of 400 cm$^{-1}$. Results obtained with different PESs
       are indicated by the symbols.
              }
         \label{Fig2}
   \end{figure}
%
   \begin{figure}
   \centering
   \includegraphics[width=8.0cm,clip]{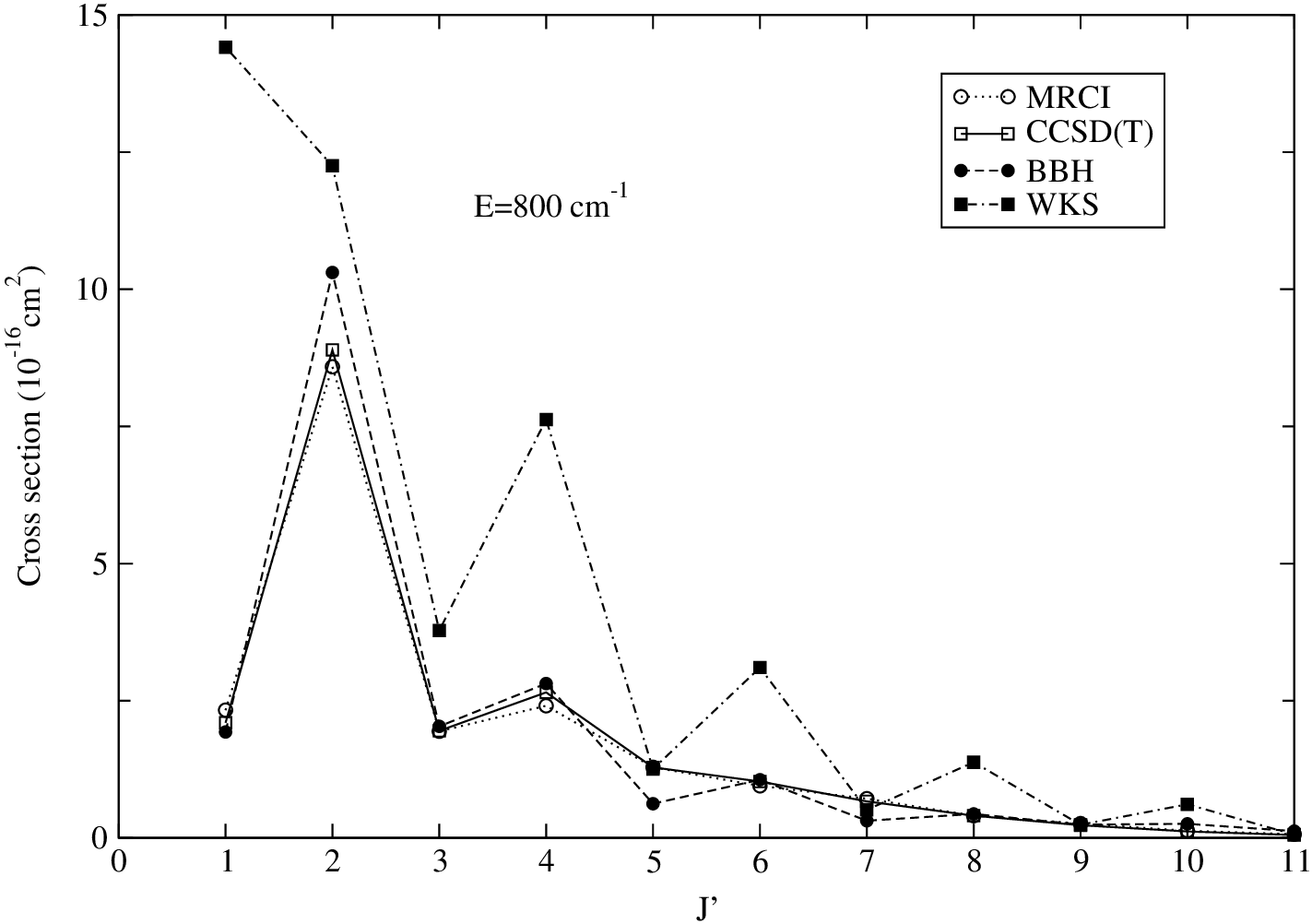}
      \caption{Same as Fig. 2, but for a center of mass
       collision energy of 800 cm$^{-1}$.
              }
         \label{Fig3}
   \end{figure}
%

\section{Conclusions}

The current scattering calculations using the new CCSD(T) and MRCI surfaces have
corroborated the expected even $\Delta J$ propensity for the H-CO
system at low collision energies. These findings indicate that the rotational
excitation rate coefficients obtained in Balakrishnan et al. (2002) using
the WKS surface are inaccurate bringing into question any astrophysical
conclusions deduced from modeling with these rate coefficients. 
Instead of being large, the $J=0\rightarrow 1$ cross section is small as
originally found by Chu \& Dalgarno (1975). The supposition 
of Green \& Thaddeus (1976) that excitation
due to molecular hydrogen and He should dominate the non-equilibrium kinetics
of CO remains valid.  
Calculations are currently underway to compute the full long-range HCO surface
at the CCSD(T) level which will allow for new large-scale scattering
calculations for an extensive range of rotational and vibrational
transitions.

\begin{acknowledgements}
We acknowledge support from NSF-CRIF grant CHE--0625237 (JMB and BCS),
from NASA grant NNG04GM59G and NSF grant AST-0607733
(PCS and BHY),
from NSF grant PHY--0555565 and NASA grant NNG06GC94G to David Huestis (NB and TJDK), 
 and from a grant from the Chemical Sciences, Geosciences, and Biosciences
Division of the Office of Basic Energy Sciences, US Department of Energy (AD and PZ).
 NB, EB, and PCS acknowledge travel
support from the Institute for Theoretical Atomic, Molecular, and Optical Physics which
is funded by the NSF.
\end{acknowledgements}

\end{document}